%% file: main.tex
\documentclass[sigconf,screen]{acmart}

\AtBeginDocument{%
  \providecommand\BibTeX{{%
    \normalfont B\kern-0.5em{\scshape i\kern-0.25em b}\kern-0.8em\TeX}}}

\copyrightyear{2022} 
\acmYear{2022} 
\setcopyright{rightsretained} 
\acmConference[CHI '22 Extended Abstracts]{CHI Conference on Human Factors in Computing Systems Extended Abstracts}{April 29-May 5, 2022}{New Orleans, LA, USA}
\acmBooktitle{CHI Conference on Human Factors in Computing Systems Extended Abstracts (CHI '22 Extended Abstracts), April 29-May 5, 2022, New Orleans, LA, USA}
\acmDOI{10.1145/3491101.3519729}
\acmISBN{978-1-4503-9156-6/22/04}

\usepackage{microtype}
\usepackage{tikz}
\usepackage{color, soul}
\usepackage{xspace}
\usepackage{xcolor}
\usepackage{verbatim}
\usepackage{enumitem}
\usepackage{etaremune}
\usepackage{soul}
\usepackage{subcaption,booktabs}
\usepackage{multirow}
\usepackage{array}

\definecolor{clightgrey}{HTML}{80868b}
\definecolor{cprompt}{HTML}{3c4043}
\definecolor{cgenback}{HTML}{F1F3F4}
\definecolor{cgenfront}{HTML}{0076BA}
\newcommand{\revised}[1]{#1\xspace}

\newcommand{\note}[2]{{\color{#1}{[#2]}}\xspace}

\newcommand{\todo}[1]{{\color{red}#1}}
\newcommand{\sherry}[1]{\note{purple}{Sherry: #1}}
\newcommand{\carrie}[1]{\note{orange}{Carrie: #1}}

\newcommand{\mike}[1]{\note{teal}{Mike: #1}}

\newcommand{\exinline}[1]{{\color{cprompt}``#1''\xspace}}
\newcommand{\quoteinline}[1]{{\color{cprompt}\emph{``#1''}\xspace}}

\newcommand{\eg}{\emph{e.g.,}\xspace}%
\newcommand{\ie}{\emph{i.e.,}\xspace}

\newcommand*\textcircle[1]{\tikz[baseline=(char.base)]{
            \node[shape=circle,draw,inner sep=0.5pt] (char) {#1};}}
\newcommand{\paragraphBold}[1]{\paragraph{\emph{\textbf{#1}}}}

\newcommand{\qbox}[2]{%
	\medskip
	\begin{center}
	    \noindent\fbox{
		\begin{minipage}{0.95\linewidth}
			\textbf{Q:} #1\newline
			\textbf{A:} #2
		\end{minipage}
	}
	\end{center}
}

\newcommand{\sysname}{PromptChainer\xspace}

\begin{document}

\title[\sysname: Chaining Large Language Model Prompts through Visual Programming]{\sysname: Chaining Large Language Model Prompts \\through Visual Programming}

\author{Tongshuang Wu}
\authornote{The work was done when the author was an intern at Google Inc.}
\authornote{Equal contribution.}
\email{wtshuang@cs.washington.edu}
\affiliation{%
  \institution{University of Washington}
  \country{USA}
}
\author{Ellen Jiang}
\authornotemark[2]
\email{ellenj@google.com}
\affiliation{%
  \institution{Google Research}
  \country{USA}
}

\author{Aaron Donsbach}
\email{donsbach@google.com}
\affiliation{%
  \institution{Google Research}
  \country{USA}
}

\author{Jeff Gray}
\email{jeffgray@google.com}
\affiliation{%
  \institution{Google Research}
  \country{USA}
}

\author{Alejandra Molina}
\email{alemolinata@google.com}
\affiliation{%
  \institution{Google Research}
  \country{USA}
}

\author{Michael Terry}
\email{michaelterry@google.com}
\affiliation{%
  \institution{Google Research}
  \country{USA}
}
\author{Carrie J. Cai}
\email{cjcai@google.com}
\affiliation{%
  \institution{Google Research}
  \country{USA}
}
\renewcommand{\shortauthors}{Wu and Jiang et al.}

\ccsdesc[500]{Human-centered computing~Empirical studies in HCI}
\ccsdesc[300]{Human-centered computing~Interactive systems and ools}
\ccsdesc[100]{Computing methodologies~Machine learning}


\input{sections/abstract}

\maketitle

\input{sections/intro}
\input{sections/relate}
\input{sections/ui}

\input{sections/user_study}
\input{sections/conclusion}

\bibliographystyle{ACM-Reference-Format}
\bibliography{ref}

\appendix
\input{sections/appendix_additional_chain}

\end{document}

%% file: sections/abstract.tex
\begin{abstract}

While LLMs have made it possible to rapidly prototype new ML functionalities, many real-world applications involve complex tasks that cannot be easily handled via a single run of an LLM. 
Recent work has found that chaining multiple LLM runs together (with the output of one step being the input to the next) can help users accomplish these more complex tasks, and in a way that is perceived to be more transparent and controllable. 
However, it remains unknown what users need when \textit{authoring} their own LLM chains -- a key step to lowering the barriers for non-AI-experts to prototype AI-infused applications. 
In this work, we explore the LLM chain authoring process.
We find from pilot studies that users need support transforming data between steps of a chain, as well as debugging the chain at multiple granularities. To address these needs, we designed \emph{\sysname}, an interactive interface for visually programming chains. 
Through case studies with four \revised{designers and developers}, we show that \sysname supports building prototypes for a range of applications, and conclude with open questions on scaling chains to even more complex tasks, as well as supporting low-fi chain prototyping.
\end{abstract}

%% file: sections/intro.tex
\section{Introduction}

Large language models (LLMs) have introduced new possibilities for prototyping with AI~\cite{yang2020re}.
Pre-trained on a large amount of text data, models like GPT-3~\cite{brown2020language} and Jurassic-1~\cite{J1WhitePaper} encode enough information to support \emph{in-context learning}:
they can be easily customized at run time (without any re-training needed) to handle new tasks, simply by taking in natural language instructions called \emph{prompts}.
For example, a user could customize a pre-trained, general purpose LLM to create an ad-hoc search engine for musicians by giving it the prompt string: \exinline{Genre: Jazz; Artist: Louis Armstrong. Genre: Country; Artist: }. An LLM would likely continue the prompt with the name of a country artist, e.g. \exinline{Garth Brooks.}
Beyond this toy example, non-ML experts have used prompting to achieve various ML functionalities in real-time, including code generation, question answering, creative writing, etc.~\cite{swanson2021story, mishra2021natural, brown2020language}.
Recent work on \textbf{prompt-based prototyping}~\cite{jiang2022prototype} found that, with LLMs' fluid adaptation to natural language prompts, non-ML experts (e.g. designers, product managers, front-end developers) can now prototype custom ML functionality with lower effort and less time, as they bypass the otherwise necessary but expensive process of collecting data and training models upfront~\cite{jiang2022prototype,bommasani2021opportunities}.

\begin{figure*}[t]
\centering

\includegraphics[trim={0 34cm 8cm 0cm}, clip, width=0.9\textwidth]{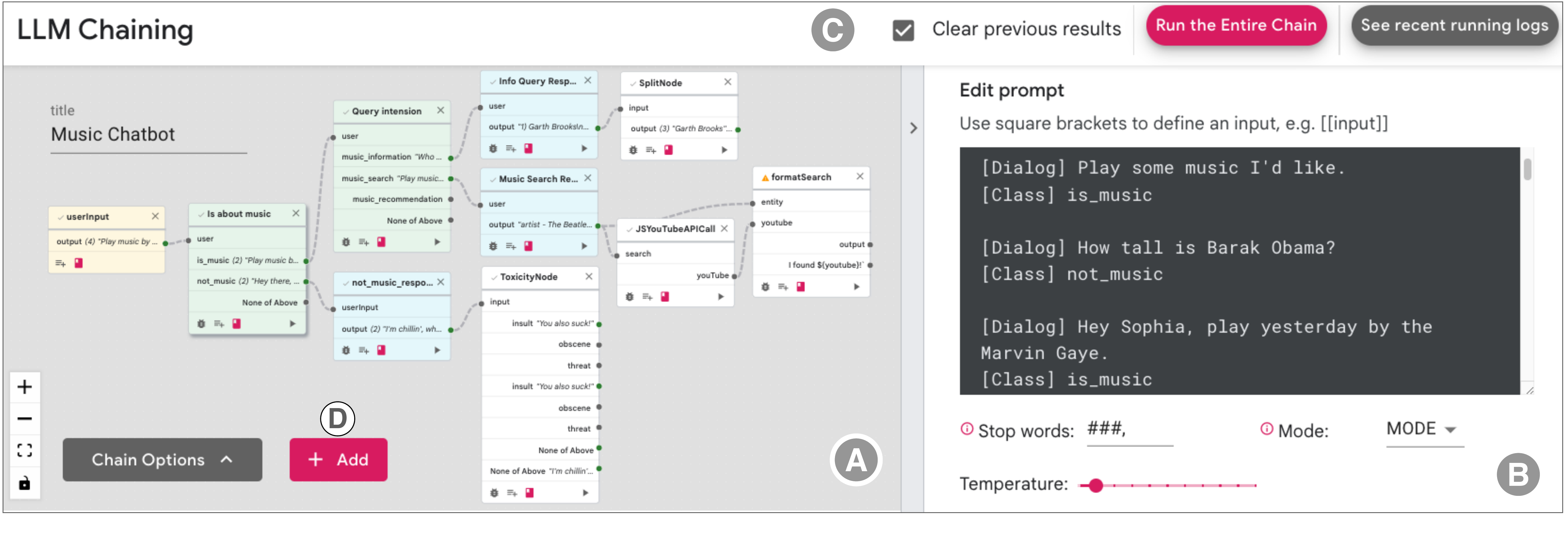}
\vspace{-10pt}
\caption{
The \sysname interface. 
(A) The \emph{Chain View} visualizes the chain structure with node-edge diagrams (enlarged in Figure~\ref{fig:example}), and allows users to edit the chain by adding, removing, or reconnecting nodes.
(B) The \emph{Node View} supports implementing, improving, and testing each individual node, \eg editing prompts for LLM nodes.
\sysname also supports running the chain end-to-end (C). 
}
\Description[The figure shows an overview of the PromptChainer interface.]{The figure shows an overview of the PromptChainer interface. On the left, we have a Chain View which visualizes the chain structure with node-edge diagrams. On the right, we have a Node View that supports implementing, improving, and testing each individual node.}
\vspace{-5pt}
\label{fig:ui}
\end{figure*}

Despite the demonstrated versatility of LLMs, many real-world applications involve complex or multi-step tasks that are nontrivial for a single run of an LLM.
For example, a music-oriented chatbot (which we build in Figure~\ref{fig:example}) may require an AI to first determine a user's query type (\eg find artists by genre as shown above, or find songs given artists, etc.), before generating a response based on the query type. 
As a result, designers and developers may struggle to prototype realistic applications with only a single LLM prompt. 
In response, \revised{we previously proposed \emph{Chaining} multiple LLM runs} together~\cite{wu2021ai}, \ie decomposing an overarching task into a series of highly targeted sub-tasks, mapping each to a distinct LLM step, and using the output from one step as an input to the next.
They observed that people can effectively \emph{use} Chaining: they could complete more complex tasks in a more transparent and controllable way.
However, it remains an open question how to support users in \emph{authoring} their own LLM chains.
For designers and developers to apply chaining to their own prototypes, they need to not only prompt within each individual LLM step, but also design the overarching task decomposition.
Such a process requires targeted tooling, akin to end-user programming~\cite{Burnett2004EnduserSE, burnett2010end}.

In this work, we examine the user experience of \emph{authoring} LLM chains.
Through formative studies, we distill three unique challenges that emerge from the extreme versatility of LLMs: 
(1) the overhead of fully utilizing LLM capabilities, 
(2) the tendency of inadvertently introducing errors to the chain when prompting, and (3) the cascading errors caused by blackbox and unstable LLM generations.
Addressing these challenges, we propose \sysname, a chain authoring interface that provides scaffolds for building a mental model of LLM's capabilities, handling arbitrary LLM data formats, defining a ``function signature" for each LLM step, and debugging cascading errors.
We conduct case studies with four designers and developers, who proposed and built chains for their own realistic application ideas (\eg chatbots, writing assistants, etc.)
Our qualitative analysis reveals patterns in how users build and debug chains: 
(1) users build chains not only for addressing LLM limitations, but also for making their prototypes extensible;
(2) some users constructed one step of a chain at a time, whereas others sketched out abstract placeholders for all steps before filling them in; 
(3) the interactions between multiple LLM prompts can be complex, requiring both local and global debugging of prompts. 

We also observed some additional open challenges, and conclude with discussion on future directions: 
First, how can we scale chains to tasks with high interdependency or logical complexity, while still preserving global context and coherence?
Second, how can we find a ``sweet spot'' for prompting such that users can quickly low-fi prototype \emph{multiple} alternative chains, without investing too much time designing any single prompt?

%% file: sections/relate.tex
\section{Background: Large Language Models, Prompting and Chaining}
A generative language model is designed to continue its input with plausible output (\eg given a prompt \exinline{I went to the}, it might auto-complete with \exinline{coffee shop}). 
However, when pre-trained on billions of samples from the Internet, recent LLMs can be adapted on-the-fly to support user-defined use cases like code generation, question answering, creative writing, etc.~\cite{brown2020language, swanson2021story}.
To invoke the desired functionalities, users need to write \emph{ prompts}~\cite{lu2021fantastically, betz2021thinking, liu2021makes} that are appropriate for the task.
The most common patterns for prompting are either zero-shot or few-shot prompts. 
Zero-shot prompts directly describe what ought to happen in a task. For example, we can enact \emph{Classification} in Figure~\ref{fig:node} with a prompt such as \exinline{Is the statement: `hey there, what's up' about music?}
In contrast, few-shot prompts show the LLM what pattern to follow by feeding it examples of desired inputs and outputs: \exinline{[Dialog] Play some music I like. [Class] is\_music [Dialog] hey there, what's up [Class]}. Given this prompt, the LLM may respond with \exinline{not\_music} (full prompt in Figure~\ref{fig:node}).

While a single LLM enables people to prototype specific ML features~\cite{jiang2022prototype}, their inherent limitations (\eg lack of multi-step reasoning capabilities) make them less capable of prototyping complex applications.
In response, \revised{we previously proposed proposed the notion of chaining multiple LLM prompts together}, and demonstrated the utilities of \emph{using} chains~\cite{wu2021ai}. 
We follow up on their work to explore how users can \emph{author} effective chains for prototyping.

%% file: sections/ui.tex
\begin{figure*}[t]
\centering
\includegraphics[trim={0 16cm 15.5cm 0cm}, clip, width=0.96\textwidth]{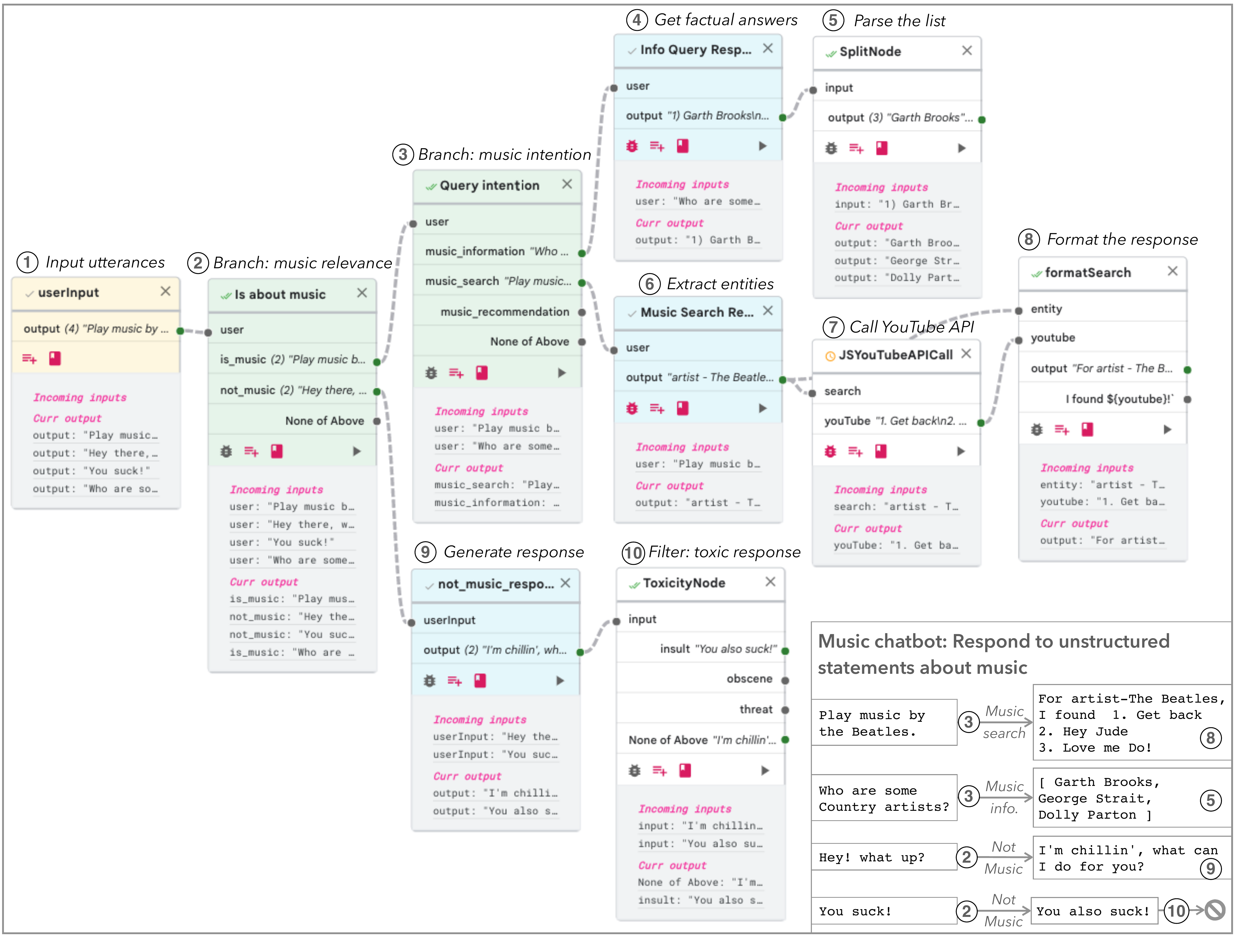}
\vspace{-16pt}
\caption{
An example chain for prototyping music chatbot, modified from a pilot user's chain (its overview is in Figure~\ref{fig:ui}).
We provide primary input-output examples, and annotate the node functionalities are annotated inline.
}
\Description{The figure an example chain for prototyping music chatbot, modified from a pilot user's chain.}

\label{fig:example}
\medskip
\input{tables/case_table}
\end{figure*}
\footnotetext{From Tensorflow.js \url{https://github.com/tensorflow/tfjs-models/tree/master/toxicity}}

\section{\sysname: Interface Requirement Analysis \& Design}

\subsection{Requirement Analysis}
\label{subsec:requirements}

\begin{figure*}[h]
\centering
\includegraphics[trim={0 36cm 20cm 0cm}, clip, width=0.85\textwidth]{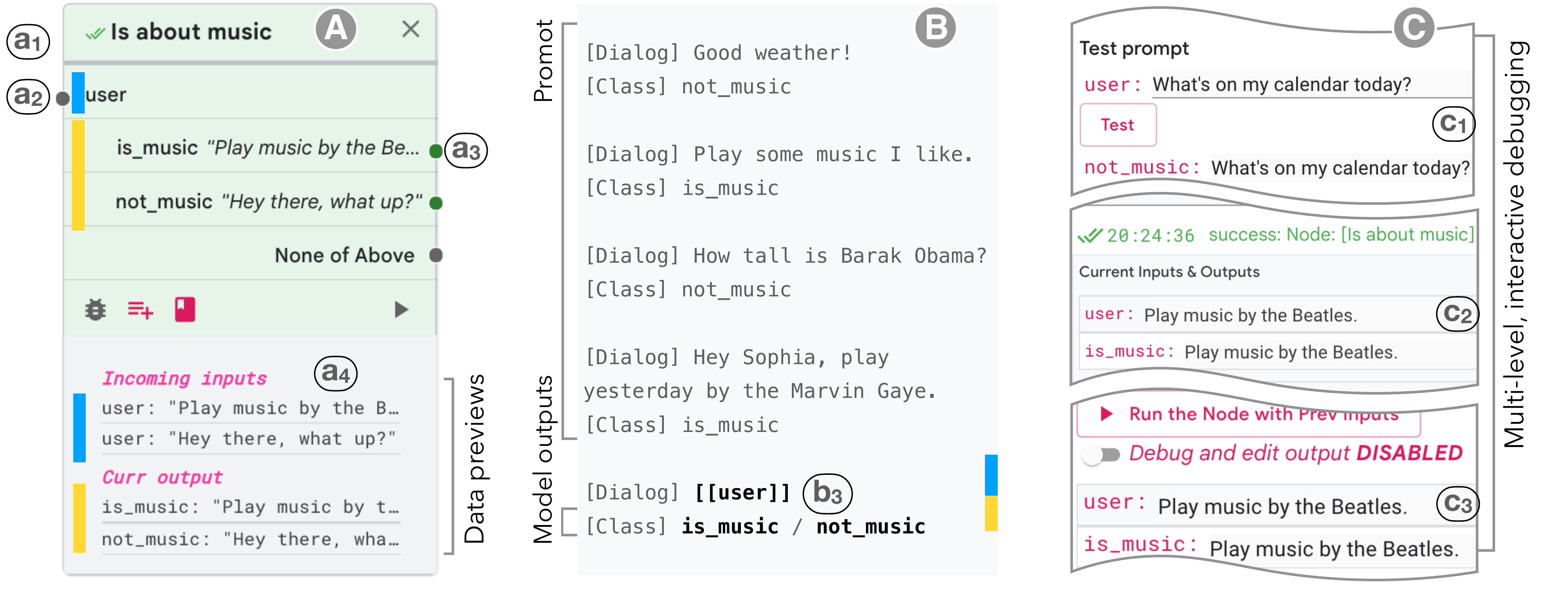}
\vspace{-15pt}
\caption{
An expansion of Figure~\ref{fig:example}, \texttt{is about music}: 
(A) Node visualization: the node has an status icon ($a_1$), a list of named input ($a_2$) and output handles ($a_3$), as well as detailed data previews ($a_4$).
(B) Implementation: the handle names are synchronized with the underlying prompt template ($b_1$).
(C) We can debug the node at multiple levels. }
\vspace{-5pt}
\label{fig:node}
\end{figure*}

To inform the design of \sysname, we conducted a series of formative studies with a team of two software engineers and three designers over the course of three months.
These formative studies included sessions reflecting on their experiences authoring LLM chains without any assistance, as well as iterative design sessions where they proposed and built their chains-of-interest in early prototypes for several rounds, reporting back their pain points (the chains from pilot studies are in Appendix~\ref{appendix:pilot_chains}).
We summarize our observations into the following authoring challenges:

\begin{enumerate}[nosep,label=C.\arabic*]



\item \textbf{The versatility of LLMs and need for data transformations}:
\label{require:model_capability}
The versatility of LLMs means that users need to develop a mental model of their capabilities. LLMs also produce outputs in arbitrary string formats, making it nontrivial to transform the output of upstream LLM steps to be compatible with the input to downstream steps.

\item \textbf{The instability of LLM function signatures}:
\label{require:connect}
LLMs also lack stable \emph{function signatures}, \ie the semantic types of their outputs easily vary with LLM prompts.
This complicates local chain iterations: for example, if a user's edits on a prompt unintentionally make an LLM step output \emph{numbered lists} instead of \emph{short phrases}, this would introduce input errors to the following steps, and thereby break the entire chain.

\item\textbf{The likelihood of cascading errors}:
\label{require:debugging}
The black-box nature of LLMs means that sub-optimal or even unsafe output in a single step could potentially lead to cascading errors across an LLM chain (see \cite{tech_debt} for a similar observation in traditional machine learning pipelines).
\end{enumerate}

\subsection{Interface Design}

We design the interface in Figure~\ref{fig:ui} in response to the challenges, with a \emph{Chain View} (Figure~\ref{fig:ui}A) for authoring the chain structure, a \emph{Node View} (Figure~\ref{fig:ui}B) for authoring a single step (node) of the chain, and support for chain debugging.

\textbf{The Chain View is the visual panel for building and viewing chains.} As in Figure~\ref{fig:ui}A, each rectangle depicts a \textbf{node}, which represents a single step in the chain, with the edges between them denoting how these nodes are \emph{connected}, or how the output of one node gets used as the input to the next.

\emph{Node visualization.}
As shown in Figure~\ref{fig:node} (a zoomed-in node of Figure~\ref{fig:example}~\textcircle{2}), each node can have one or more named inputs ($a_2$) and outputs ($a_3$), which are used to connect nodes.
Inspired by several existing node-edge-based visual programming platforms\footnote{\eg Maya: \url{https://www.autodesk.com/products/maya/overview}; Node-RED: \url{https://nodered.org/}}, we provide node previews to increase chaining transparency, including a status icon highlighting whether the node contains errors (Figure~\ref{fig:node}$a_1$), as well as inline and detailed data views ($a_3$ and $a_4$).

\emph{Node Types.} 
As summarized in Figure~\ref{table:node_types}, we define several types of nodes to cover diverse user needs.
At its core are two types of \textbf{LLM nodes}: Generic LLM nodes and LLM Classifier nodes (See the node definitions and examples in Figure~\ref{table:node_types}). Users can implement these nodes by providing a natural language prompt, call an LLM with the prompt as input, and use the LLM outputs accordingly.
\sysname also provides \textbf{helper nodes} that address common data transformation (\ref{require:model_capability} in Section~\ref{subsec:requirements}) and evaluation needs (\ref{require:debugging}), or to allow users to implement their own custom JavaScript (JS) nodes.
Finally, to support users in prototyping AI-infused applications, \sysname provides \textbf{communication nodes} for exchanging data with the external world (\eg external API calls). 

\emph{Example gallery.} 
To address the versatility challenge of LLMs (\ref{require:model_capability}),
\sysname provides examples of frequently composed (sub-)chains, to help users develop a mental model of which capabilities are likely to be useful. 
These examples also serve as a soft nudge towards a set of prompting patterns, such that users' prompts are more likely to be compatible with predefined processing nodes. 
For example, Figure~\ref{fig:example} \textcircle{4} is forked from an \emph{Ideation} example that returns numbered lists \exinline{1) Garth Brooks 2) George Strait...}, which is parsable with the provided \textcircle{5} \texttt{Split by number} node.


\textbf{The Node View allows users to inspect, implement, and test individual nodes} (Figure~\ref{fig:ui}B).
When a node is selected, the panel changes in accordance with the Node Type.
\sysname automatically parses the input names of a node based on the LLM prompt for that node (or, for JavaScript helper nodes, based on the function signature).
For example, in Figure~\ref{fig:node}, the input handle $a_1$ ``\texttt{user}'' is synchronized with the bolded placeholder string \texttt{[[user]]} ($b_1$) in its corresponding prompt template, meaning that the input to $a_1$ will be used to fill in $b_1$ in the prompt.
If a user changes $b_1$ to \eg \texttt{[[sentence]]}, $a_1$ would get renamed to ``\texttt{sentence},'' such that there will be no outdated handles.
As such, PromptChainer automatically updates the global chain to be consistent with users' local edits (addressing \ref{require:connect}).


\textbf{Interactive debugging functionalities.}
To address the cascading error challenge (\ref{require:debugging}), \sysname supports chain debugging at various levels of granularity:
First, to \emph{unit test} each node, users can use the provided \emph{testing block} (Figure~\ref{fig:node}$c_1$) to test each node, with examples independent of the remaining chain.
Second, to perform \emph{end-to-end assessment}, users can run the entire chain and \emph{log the outputs per node}, such that the ultimate chain output is easy to retrieve (Figure~\ref{fig:node}$c_2$).
Third, to help users map global errors to local causes, \sysname supports \emph{breakpoint debugging} (Figure~\ref{fig:node}$c_3$), and allows users to directly edit the output of a node before it is fed into the next node. 
By fixing intermediate node outputs, users can test a subset of downstream nodes independent of earlier errors.

%% file: tables/case_table.tex
\renewcommand{\arraystretch}{0.8}
\fontsize{7.5}{8}\selectfont
\begin{tabular}{p{0.005\linewidth} p{0.135\linewidth} | p{0.55\linewidth} | p{0.16\linewidth}}
\toprule
\multicolumn{2}{l}{\textbf{Node Type}} & \textbf{Description} & \textbf{Example in Figure~\ref{fig:example}} \\

\midrule\midrule
\multirow{2}{*}{\rotatebox[origin=c]{90}{LLM}}
& Generic LLM 
    & Use the LLM output directly as the node output. 
    & \textcircle{4} \textcircle{6} \textcircle{9} \\
\cmidrule(lr){2-4}
& LLM Classifier 
    & Use LLM output to filter and branch out inputs. 
    & \textcircle{2} \textcircle{3} \\

\midrule\midrule
\multirow{3}{*}{\rotatebox[origin=c]{90}{Helper}} 
& Evaluation 
    & Filter or re-rank LLM outputs by human-designed criteria, \eg politeness.
    & \textcircle{10} Toxicity classifier\footnotemark\\
\cmidrule(lr){2-4}
& Processing 
    & Pre-implemented JavaScript functions for typical data transformation. 
    & \textcircle{5} Split by number\\
\cmidrule(lr){2-4}
& Generic JavaScript 
    & User-defined JS functions, in case pre-defined helpers are insufficient.
    & \textcircle{8} Format the query\\

\midrule\midrule
\multirow{3}{*}{\rotatebox[origin=c]{90}{Comm.}} 
& Data Input 
    & Define the input to a chain.
    & \textcircle{1}\\
\cmidrule(lr){2-4}
& User Action 
    & Enables external (end user) editing on intermediate data points.
    & (Figure~\ref{fig:user_chains} \textcircle{11})\\
\cmidrule(lr){2-4}
& API Call 
    & Call external functions to connect professional services with LLMs.
    & \textcircle{6} Call YouTube API\\
\bottomrule
\end{tabular}
\vspace{-10pt}
\caption{A summary of node types, including the core \textbf{LLM nodes}, \textbf{helper nodes} for data transformation and evaluation, and \textbf{communication nodes} for exchanging LLM data with external users or services.}
\vspace{-5pt}
\Description{The table summarizes all the node types in detail. It includes the core LLM nodes, helper nodes for data transformation and evaluation, and communication nodes for exchanging LLM data with external users or services.}
\label{table:node_types}


%% file: sections/user_study.tex
\section{User Feedback Sessions}

We conducted a preliminary study to understand what kinds of chains would users want to build, the extent to which \sysname supports their needs, and what additional challenges users face.

\subsection{Study design}
Because Chain authoring goes beyond single prompts to chaining multiple prompts together, we recruited four participants  (3 designers, 1 developer \revised{within Google}) who had at least some prior experience with \revised{non-chained} prompts: P1 and P2 had prior experience writing prompts, and P3 had seen some LLM demos.
\revised{We personally reached out to these participants in different product teams, to prioritize interests and experience in a wide range of domains.}
\revised{Before the study session, participants spent 30 minutes on preparation}: They watched a 10-minute tutorial on interface features.
They were also asked to prepare a task beforehand that they believed would require multiple runs of the LLM, envision the LLM call for each step, and draft each of those prompts.
This way, the study could focus on chaining prompts together rather than writing the initial prompts.
In the hour-long actual study, participants loaded their prompts and authored their envisioned Chain while thinking aloud.
To observe the extent to which \sysname could support iteration, we asked participants to describe deficiencies (if any) in their Chain, and modify their Chains. 
We observed and recorded their entire task completion sessions, and later transcribed their comments for qualitative analysis.
In total, participant spent approximately 90 minutes, and received a \$75 gift credit for their time.

\revised{\textbf{Underlying LLM.} 
All of our experiments (including pilot study) rely on the same underlying LLM called LaMDA~\cite{thoppilan2022lamda}\footnote{We used a non-dialog version of the model.}: a 137 billion parameter, general-purpose language model.
This model is roughly equivalent to the GPT-3 model in terms of size and capability: it is trained with more than 1.5T words of text data, in an auto-regressive manner using a decoder-only Transformer structure.
It has comparable performances with GPT-3 on a variety of tasks, and behaves similarly in its ability to follow prompts.}

\begin{figure*}[t]
\centering
\includegraphics[trim={0 19.5cm 18cm 0cm}, clip, width=1\textwidth]{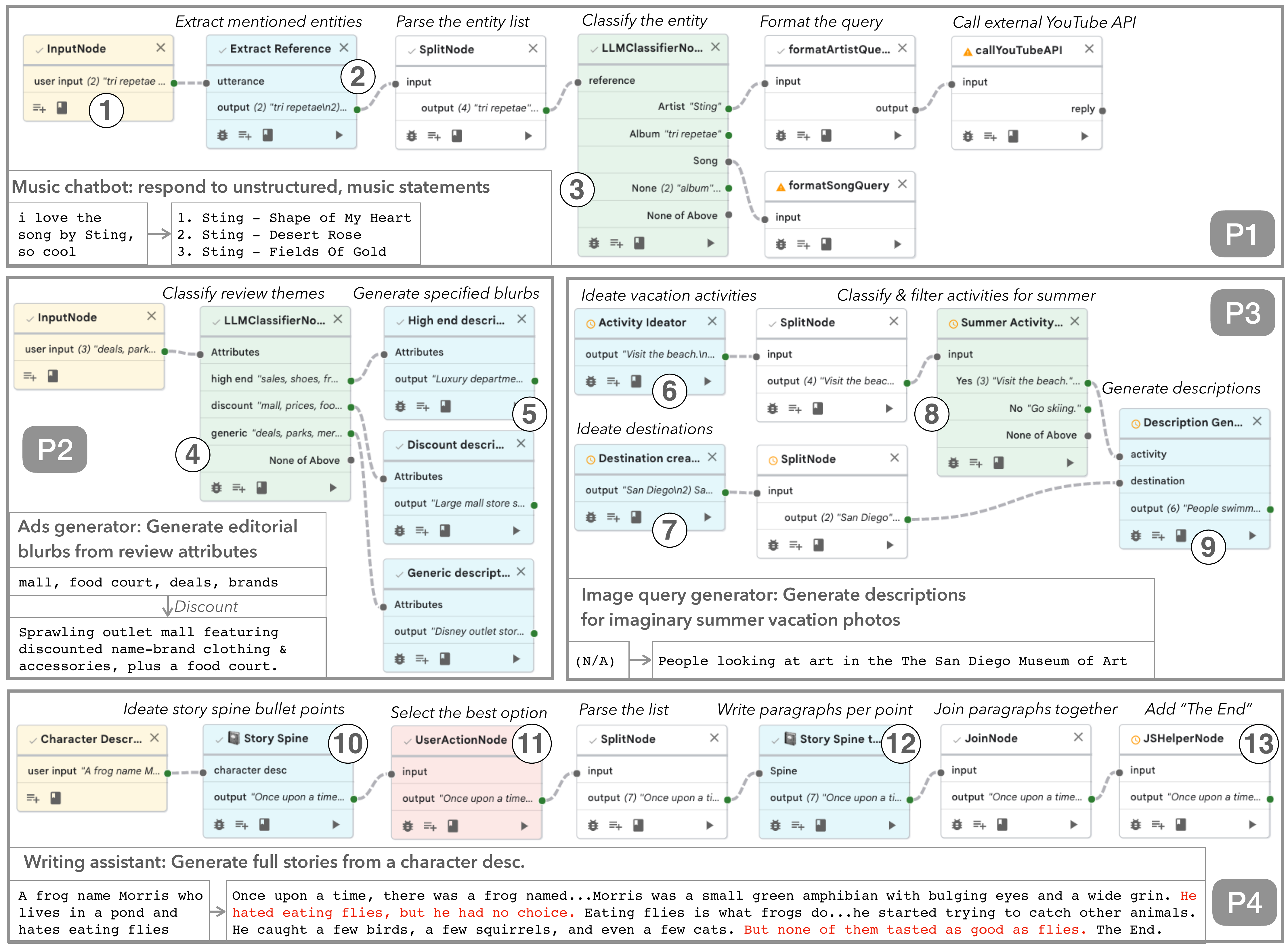}
\vspace{-20pt}
\caption{
Four different chains built by user study participants. P1 and P2's chains used parallel branching logic , whereas P3 and P4's chains depict iterative content processing. 
The full details are in Figure~\ref{fig:user_chains_full}, Appendix~\ref{appendix:pilot_chains}.
}
\vspace{-10pt}
\label{fig:user_chains}
\end{figure*}

\subsection{Study Results}

We analyze our study to answer the three questions listed below.

\vspace{-5pt}
\qbox
{What kinds of chains would users want to build?}
{Users proposed diverse tasks, some that used branching logic, and some that iterated on content. They used chaining not only to address single-prompt limitations, but also to make their prototypes \textit{extensible}.}

%
\textbf{Chaining patterns.}
All participants successfully built their desired chains during the study session, with the chains containing on average $5.5\pm 0.9$ nodes.
As shown in Figure~\ref{fig:user_chains}, participants built chains for various kinds of tasks, ranging from music chatbot, to ads generator, to writing assistants.
Their chain structures reflect different high-level patterns:
(1) P1 and P2 built chains with \textbf{parallel logic branches}, similar to decision trees~\cite{geng2018decision}.
For example, P2's chain sought to generate specialized descriptions for different kinds of product review attributes. 
They first classified whether attributes were ``high end'', ``discount'', or ``generic'' with \textcircle{4}, which determined which downstream node (specialized LLM description generator) would run.
(2) P3 and P4 built chains that \textbf{incrementally iterate on content}. These chains usually take a divide-and-conquer strategy~\cite{kittur2011crowdforge}.
For example, P4 wrote one story by first generating a list of story outlines \textcircle{10}, then generating paragraphs per point \textcircle{12}, and finally merging them back \textcircle{13}.


\textbf{Chaining rationales.}
While we primarily expected participants to author chains for the purpose of combating LLM limitations (as we previously observed~\cite{wu2021ai}), interestingly, 
participants also inserted chaining steps to \textbf{make their prototypes more generalizable}.
For example, though P3 knew they could directly build an ideation node for ``summer vacation activities'', they instead chose to build a more general ideator \textcircle{6} combined with a summer activity classifier \textcircle{7}, such that they could \quoteinline{switch the classifier for any other time or variables, for variability and flexibility.}
Further, P4 mentioned the desire for \textbf{taking intermediate control}.
Because he noticed the top-1 LLM generations were not necessarily their favorite, he built a chain to make active interventions: He first \emph{over-generated} three candidate story spines in \textcircle{10}, from which he would select and refine one of them in \textcircle{11} for subsequent paragraph expansions.

\qbox
{To what extent does \sysname support users in iteratively authoring and improving their chains?}
{\sysname supported a variety of chain construction strategies and enabled multi-level debugging.}

\paragraphBold{Chain construction.}
Participants applied various chain construction strategies.
P1 performed a top-down approach, \ie they connected blank, placeholder nodes first to illustrate the imaginary task decomposition before filling in the prompts. 
In contrast, the other three participants worked on each node one at a time, before moving on to the next node: whereas P2 carefully ran and tested each node, others created ``rough drafts'', starting with basic prompts and deferring detailed refinement of prompts until after a draft chain was finished (P3: \quoteinline{I should probably move on with this, I want to fine-tune my LLM prompt later.}). These varied chain construction strategies indicate that \sysname can support different pathways for chain prototyping, but that a common user tendency may be to work one node at a time.
%
%

To further characterize the node utilities, we analyzed the node distributions in both the user study chains and those from pilot users (8 in total).
We found that \emph{pre-defined helpers could cover most of the chaining needs}: participants used three times as many pre-defined helpers (13 in total) as customized JS nodes (4).
One author further codified all the LLM nodes according to the primitive LLM operations previously identified~\cite{wu2021ai}, and found that out of the 27 LLM nodes, 7 were for categorizing inputs, 13 for sourcing information from the LLM, and 7 for re-organizing the input.
This variety in utilization may have resulted from the \sysname's example galleries.
For example, P4 successfully created their own LLM classifier by forking a simple default example, even though they were less familiar with prompting.


\textbf{Chain debugging.}
When they completed constructing their chains, all participants ran the chain end-to-end (Figure~\ref{fig:node}$C_2$) as an \exinline{initial debugging strategy} (P1 and P4).
Afterwards, they usually attributed chain failure to particular LLM nodes (P1: \quoteinline{easy to pinpoint unexpected outputs from the data preview}), and performed local debugging. 
P1 appreciated the \emph{breakpoint functionality} (Figure~\ref{fig:node}$C_3$), as they did not need to take the chain apart in order to debug one of the nodes;
P3, on the other hand, relied on the independent \emph{testing block} (Figure~\ref{fig:node}$c_1$) when debugging the \texttt{Description Generator} \textcircle{9}, as a way to avoid expensive executions on multiple inputs coming from prior nodes.

Interestingly, most participants made some non-trivial refinements to their pre-built prompts \emph{in the interface}, even though they had spent a fair amount of time doing prompt engineering before the study.
We hypothesize that being able to observe the interaction effects \emph{between} nodes affected their understanding and expectations of each local node.
For example, when constructing the story creation chain, P4 wanted to add a final ending, \exinline{The End}, to the generated story.
They first tried to always generate \exinline{The End} as the final bulletpoint in the story outline (``Story Spine") in \textcircle{10}, but realized that this would cause paragraph generator \textcircle{12} to produce a paragraph repeating \exinline{The End The End The End.}
They therefore removed this generation from \textcircle{10}, and instead (with some help from the study facilitator) made a final JavaScript helper node \textcircle{13} for appending the text \exinline{The End}.
This suggests that \sysname can help users \emph{discover} errors, though future research is needed in supporting them to \emph{resolve} identified problems through alternative solutions.

\qbox
{What are remaining challenges in chain authoring?}
{Ensuring coherence between interdependent sub-tasks; tracking chains with complex logic.}

\textbf{Chains with interdependent parallel tasks can lead to decreased coherence.}
Because P4's story writing chain independently generated a paragraph for each point in the outline, the final essay lacked coherence: though the first several sentences followed the input description (\exinline{Morris...hates eating flies}), the final sentence instead hinted that Morris \emph{likes} flies (\exinline{none of them tasted as good as flies}).
One pilot study user faced a similar challenge, and created another input node to manually track previous outputs. In the future, it may be worthwhile to further investigate methods that consider inter-dependency between parallel sub-tasks~\cite{swanson2021story}.

\textbf{Chains that involve more complex decomposition can be overwhelming to track.}
In P1's music chatbot chain, the extractor node \textcircle{2} produces \emph{a list of} candidate entities per input. Thus, it became unclear how the entities fed into the classifier \textcircle{3} mapped to the original input node \textcircle{1}.
We hope to enhance the tracing capabilities of \sysname. 
For example, future work could enable customized chain grouping: 
Instead of running one or all of the nodes, participants can explicitly select a subset to run.
We may also add execution visualizations (\eg taking inspiration from Python Tutor\footnote{\url{https://pythontutor.com/}}), to highlight the mapping from the original input all the way to the final output. 

\subsection{Discussion and Limitations}
\label{subsec:discuss}

Because participants were asked to pre-create some LLM prompts for their desired sub-tasks prior to the study, this may have unintentionally led to participants feeling invested in their prompts and their particular chain decomposition, making them less inclined to consider other chain structures or scrap the prompts they had already created.
Yet, prior work in prototyping indicates that concurrently considering multiple alternatives (\eg parallel prototyping~\cite{dow2010parallel}) can lead to better outcomes.
Thus, future work could explore how to encourage low-fi prototyping of \emph{multiple} possible chains: in other words, how can users create half-baked prompts for each step, such that the feasibility of an entire chain can be rapidly tested, without investing too much time designing each prompt? 
For example, \sysname could perhaps encourage users to start with only one or two examples in a few-shot prompt, or start with a very simple zero-shot prompt  (even if they don't initially perform reliably) to reduce initial time invested in each prompt. 

Given time constraints in the study, users may have also picked tasks that were naturally easy to decompose. In the future, it would be worthwhile to explore task decomposition strategies for even larger and more complex tasks. For example, \sysname could help encourage users to further decompose a node in the chain into more nodes, if extensive prompting efforts appear unsuccessful. 

%% file: sections/conclusion.tex
\section{Conclusion}
We identified three unique challenges for LLM chain authoring, brought on by the highly versatile and open-ended capabilities of LLMs. 
We designed PromptChainer, and found that it helped users transform intermediate LLM output, as well as debug the chain when LLM steps had interacting effects.
Our study also revealed interesting future directions, including supporting more complex chains, as well as more explicitly supporting ``half-baked" chain construction, so that users can easily sketch out a chain structure without investing too much time prompting upfront.

%% file: sections/appendix_additional_chain.tex
\begin{figure*}[h]
\centering

\includegraphics[trim={0 4cm 17.5cm 0cm}, clip, width=0.88\textwidth]{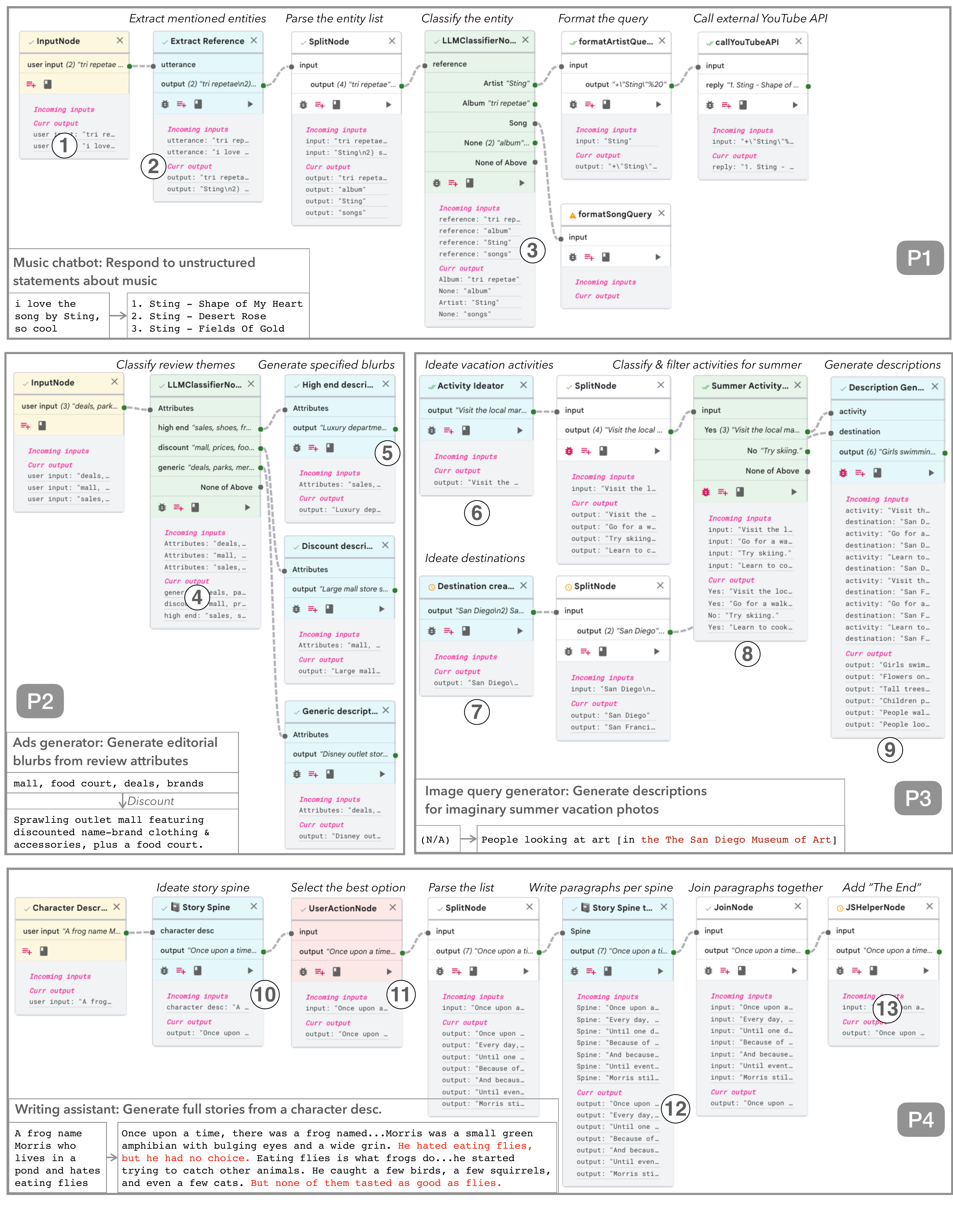}
\vspace{-20pt}
\caption{
The full details of user study chains.
}
\label{fig:user_chains_full}
\end{figure*}

\section{Sample Chains from Pilot Users}
\label{appendix:pilot_chains}

\begin{figure*}
\centering

\includegraphics[trim={0 2cm 30cm 0cm}, clip, width=0.7\textwidth]{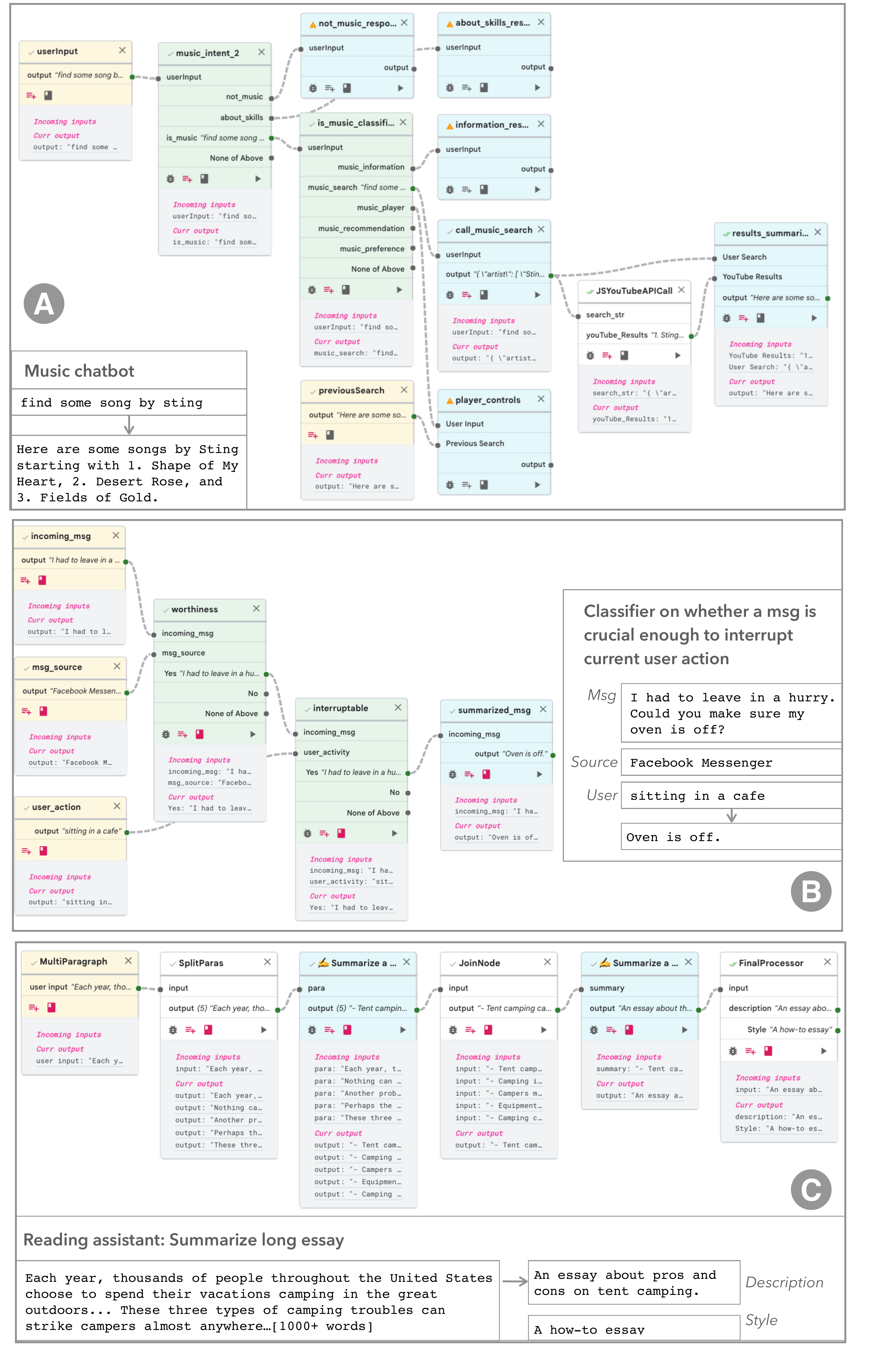}
\vspace{-15pt}
\caption{
The chains built by pilot users.
}
\label{fig:pilot_chains}
\end{figure*}
